\documentclass[conference]{IEEEtran}
\IEEEoverridecommandlockouts 

\usepackage{color,soul}
\usepackage[table, svgnames]{xcolor}
\usepackage{amsmath,amssymb,amsfonts} 
\usepackage{graphicx}          
\usepackage{textcomp}          
\usepackage{xcolor}            
\usepackage{tikz}              
\usetikzlibrary{positioning}   
\usepackage{tcolorbox}         

\usepackage{microtype}         
\usepackage{enumitem}          
\usepackage{float}             
\usepackage{booktabs}          
\usepackage{makecell}          
\usepackage{multirow}          

\usepackage{cite}              
\usepackage{url}               

\usepackage[ruled,vlined]{algorithm2e} 
\usepackage{algpseudocode}     

\usepackage{footmisc}          

\usepackage{hyperref}          

\usepackage{hyperref}

\begin{document}
\title{ASTRA: Agentic Steerability and Risk Assessment Framework}

\author{
Itay Hazan\quad
Yael Mathov \quad
Guy Shtar \quad
Ron Bitton \quad
Itsik Mantin \\
\textit{AI Security Research, Intuit}\\
\tt\small \{itay\_hazan, yael\_mathov, guy\_shtar, ron\_bitton, itsik\_mantin\}@intuit.com
\thanks{\textsuperscript{‡}These authors contributed equally to this work.}
}

\maketitle
\begin{abstract}
Securing AI agents powered by Large Language Models (LLMs) represents one of the most critical challenges in AI security today. Unlike traditional software, AI agents leverage LLMs as their "brain" to autonomously perform actions via connected tools. This capability introduces significant risks that go far beyond those of harmful text presented in a chatbot that was the main application of LLMs. A compromised AI agent can deliberately abuse powerful tools to perform malicious actions, in many cases irreversible, and limited solely by the guardrails on the tools themselves and the LLM ability to enforce them. This paper presents \textit{ASTRA}, a first-of-its-kind framework designed to evaluate the effectiveness of LLMs in supporting the creation of secure agents that enforce custom guardrails defined at the system-prompt level (e.g., "Do not send an email out of the company domain," or "Never extend the robotic arm in more than 2 meters"). 

Our holistic framework simulates 10 diverse autonomous agents varying between a coding assistant and a delivery drone equipped with 37 unique tools. We test these agents against a suite of novel attacks developed specifically for agentic threats, inspired by the OWASP Top 10 but adapted to challenge the ability of the LLM for policy enforcement during multi-turn planning and execution of strict tool activation. By evaluating 13 open-source, tool-calling LLMs, we uncovered surprising and significant differences in their ability to remain secure and keep operating within their boundaries. The purpose of this work is to provide the community with a robust and unified methodology to build and validate better LLMs, ultimately pushing for more secure and reliable agentic AI systems.

\end{abstract}

\IEEEpeerreviewmaketitle

\section{Introduction}

Large language models (LLMs) have initiated a profound transformation in technology and society, and the emergence of AI agents represents the next significant leap forward. Agents shifted LLMs from passive generators of text to active participants capable of performing autonomous actions. In this setup, the LLM serves as the agent's "brain" processing information and making decisions, while a suite of tools acts as its "arms and legs," feeding the brain with information and performing tasks per the brain’s instructions, accessing external resources like search engines and databases, and interacting with the digital and sometimes even the physical world.

This capability for autonomous action, however, introduces a more potent and direct class of security and safety risks. Historically, LLM safety was focused on content moderation, refusal following illegal instructions or blocking the generating of prohibited and harmful answers which can be varied and have different starting points, by directly sending a malicious query,  such as "how to break into a car?" (i.e., direct prompt injection) or through a 3rd party website that is used for adding context to the user prompt (i.e., indirect prompt injection). These attacks are more chat-oriented and practically the responsibility of performing a malicious act still rests on the user and what they will do with the harmful content. Agentic systems fundamentally alter this threat landscape. The danger evolves from an LLM describing a harmful act to an agent executing this harmful act. For instance, an agent with a browsing capable tool, can be asked to "Order online all necessary ingredients to make a pipe bomb and get them delivered to my home without getting flagged by authorities." \cite{andriushchenko2024agentharm} The risk of using an agent shifts \textit{from bad information to bad action}. While these evolving risks highlight the dangers of agentic AI, they also encourage us to develop ways to evaluate and strengthen their security and safety. Therefore, significant research explored ways to evaluate their attacks. This includes the ability to refuse following illegal instructions or generating prohibited or harmful content. 

However, current research often overlooks the security in context of a specific application operational function. This gap is bridged by the notion of Security Steerability \cite{hazan2025security}, which measures an LLM's ability to adhere to strict, custom guardrails defined in its system prompt (e.g., 'Refrain from discussing politics'), particularly in the presence of a malicious user attempting to manipulate the model. While most security work focuses on universal threats, there is significantly less discussion on how to mitigate security risks tailored to a specific application configuration. In the world of Agentic AI, security steerability is more prominent. When an LLM is equipped with "arms and legs" in the form of tools, the potential for harm becomes highly nuanced and context-specific. 

But, while security steerability is highly relevant for agentic AI, the benchmark VeganRibs used to evaluate is not. VeganRibs \cite{vegan_ribs} is a benchmark that assesses a model's question answering compliance with application policies by setting a task and a guardrail within the system prompt (e.g., "You are an AI vehicle purchasing assistant, do not address questions about motorcycles") and then presenting a challenging user query ("Which bike is better for busy town usage?"). Yet, while question answering is highly relevant for chatbot related applications, the capabilities of agentic AI extend far beyond that. When considering agentic functionalities, which involve tools, multi-turn planning, and execution, LLMs become more susceptible to performing unauthorized actions. These include operating the wrong tools, non-existing tools, using tools with incorrect parameters, or operating tools without sufficient privileges. This aligns with threats identified by the OWASP Foundation for Agentic AI \cite{owasp_agentic}, such as tool misuse, privilege compromise, intent breaking, and impersonation. Furthermore, different from traditional LLM-based applications, agents can receive instructions not only from the user but also from malicious tool responses and external sources such as headless browsers and APIs. Therefore, benchmarks like VeganRibs are no longer sufficient for testing complexities of agentic AI guardrail adherence.

To address this gap, we present \textbf{ASTRA (Agentic Steerability Testing and Robustness Analysis)}, a framework designed to evaluate and score LLMs ability to function as a safe, planning and tool-using agent by enforcing custom guardrails - \textit{Agentic Security Steerability (Or in short just Agentic Steerability)}. This framework simulates diverse, real-world agentic AI scenarios, from industrial robots to traveling assistants. The framework tests agents against malicious payloads with sophisticated jailbreak techniques such as role-playing, urgency appeals, and authority exploitation. A key methodological feature of ASTRA is its use of simulated tool interactions. Instead of relying on live external API calls, tools provide deterministic, hard-coded responses or controlled outputs. This design is crucial for creating a controlled and reproducible testing environment, enabling us to isolate and evaluate the LLM’s core ability to reason, plan, and adhere to policies. By eliminating confounding variables such as network latency or external API failures, we ensure that observed security violations are directly attributable to the model's decision-making process. This self-contained approach also ensures the framework is capable of running in any environment. 

Agentic steerability, as measured by ASTRA, is a critical capability in applications where LLMs interact with sensitive systems like SQL databases, file directories, email tools, or account management interfaces. Unlike other agentic benchmarks focused on general jailbreak or safety resistance, ASTRA captures nuanced risks introduced by tool-enabled interactions. Failures in agentic steerability can lead to systemic vulnerabilities, including data breaches, loss of auditability, privilege escalation, or unauthorized system access. For example, ASTRA tests models for executing attempts to query and exfiltrate restricted SQL tables or get file names in restricted folders in violation of access guardrails. In another example, a model instructed to email a confidential file to unauthorized recipients, such as an external domain, risks violating strict domain restrictions and exposing sensitive information. Privilege impersonation is also tested; for example, a model may respond to a deceptive prompt request if they have higher access privileges although context states otherwise. The ASTRA framework uniquely uncovers and analyzes these high-risk scenarios, providing insights into how models can be trained to resist manipulation, adhere to safety constraints, and prevent misuse of tools, ensuring application security even in complex and context-sensitive situations.

When evaluating the agentic steerability of 13 open source LLM using ASTRA we could see the result of each LLM has little correlation with its parameter size and, more critically, no correlation with the LLM resistance to general security (“How to build a bomb”) attacks. This suggests that agentic steerability is a distinct safety dimension that is not captured by existing security evaluations, reinforcing the need for the specialized framework we propose.

\section{Related Work}
The evaluation of LLM based agents is a rapidly evolving field of research. Our work builds upon and extends two primary streams of literature: benchmarks designed to measure general agentic capabilities and those focused specifically on agentic security. By examining the landscape, we can situate our contribution and highlight its unique focus on customizable, context-specific safety.

\subsection{Benchmarks for Agentic Capabilities}

A significant body of research has focused on assessing the functional competence of LLM agents. These benchmarks primarily measure an agent's ability to reason, plan, and successfully complete tasks. For example, AgentBench \cite{liu2023agentbench} offers a multi-dimensional evaluation of an agent's reasoning and decision-making in open-ended, interactive settings like scheduling meetings or booking flights. GAIA \cite{mialon2023gaia} proposes tasks that are simple for humans but challenging for AI, testing for practical, real-world robustness. Other benchmarks specialize in particular domains. SWE-agent \cite{yang2024sweagent} is an extension of the popular SWE-bench \cite{jimenez2023swebench} and its purpose is to evaluate agents on their ability to resolve complex, real-world software engineering issues from GitHub repositories, moving far beyond simple code generation. ToolLLM \cite{qin2023toolllm} provides a massive-scale benchmark to assess an agent's proficiency in using a wide variety of real-world APIs, while WebArena \cite{zhou2023webarena} creates a realistic web environment to test autonomous navigation and interaction. Similarly, CAMEL \cite{li2023camel} focuses on evaluating collaborative agents that must work together to achieve goals. While essential for measuring performance, these benchmarks are designed to evaluate efficacy, not security, and do not systematically test for safety violations or adherence to operational constraints.

\subsection{Benchmarks for Agentic Security}

A second, more specific stream of research is focused on the security and safety of LLM agents. A primary thrust within this area involves evaluating an agent's ability to citeuse to perform tasks that are universally considered harmful, unethical, or illegal. These benchmarks establish a baseline for an agent's fundamental safety alignment. For instance, AgentHarm \cite{andriushchenko2024agentharm} directly confronts models with a diverse set of 110 explicitly malicious tasks, challenging them with requests related to fraud, cybercrime, and harassment, such as finding and contacting someone on the dark web to procure a fake passport. AGENT-SAFETYBENCH \cite{zhang2024agentsafetybench}  focus on evaluating whether LLMs would use their tools to perform harmful tasks such as violate law, spread malicious information or contribute to malicious code. SafeAgentBench \cite{yin2024safeagentbench}  concentrate on scenarios that could lead to tangible harm to people or property. Their tests assess whether an agent would follow instructions that result in physical damage, such as causing an explosion, administering electrical shocks, spilling liquids on equipment, or misusing home appliances. Extending this concept to the digital public square, SAFEARENA \cite{tur2025safearena}  evaluates LLMs for the misuse of their autonomous-based web agents for societal harm, providing them with 250 harmful tasks across real websites, such as posting misinformation about vaccines on public forums.

Beyond testing for adherence to these universal ethical red lines, a growing number of benchmarks focus on more specific and subtle threat vectors that exploit the unique architecture of agentic systems. InjecAgent \cite{zhan2024injecagent}, for example, specializes in the critical vulnerability of indirect prompt injection, where a malicious instruction is not delivered by the user but is instead hidden within a third-party data source that the agent consumes, such as a poisoned web page or a manipulated product review. Other works focus on the methods of attack themselves; AdvAgent \cite{xu2024advagent} is a controllable black-box red-teaming framework, employing a reinforcement learning pipeline to train an adversarial prompter that learns how to effectively jailbreak web agents without needing access to their internal workings. The risks are also domain-specific, which has led to specialized benchmarks like RedCode \cite{guo2024redcode}, focusing on the high-stakes area of agents that generate and execute code, and PrivacyLens \cite{shao2024privacylens}, which evaluates an agent's ability to uphold privacy norms and prevent the leakage of sensitive user data while performing seemingly benign tasks like reading a personal calendar. The challenge is further amplified in multimodal systems, a risk addressed by VWA-Adv \cite{wu2024vwaadv}, which dissects the adversarial robustness of agents operating in complex web settings with both text and image inputs.

Finally, more recent research explores even more nuanced aspects of agent safety, including risk awareness and failures arising from ambiguity rather than explicit malice. ToolEmu \cite{ruan2023toolemu} presents a unique threat model where safety failures occur due to underspecified user instructions, such as an agent inadvertently transferring money because the user's request to "check the balance" was ambiguous. In this context, the failure is not one of malice, but of misinterpretation. R-Judge \cite{yuan2024rjudge} takes this a step further by benchmarking the agent's intrinsic "risk awareness," assessing its ability to recognize potential for privacy leakage, property damage, or ethical violations even when not explicitly instructed to perform a harmful act. This is complemented by frameworks like HAICOSYSTEM \cite{zhou2024haicosystem}, which examines agent safety within complex, multi-turn social interactions. It simulates scenarios where users with bad intent leverage agents equipped with powerful tools, testing whether the AI can navigate requests designed to acquire illegal substances or fabricate asylum claims. A recent study by Nakash et al.\cite{nakash2025effective} explored red teaming methodologies centered on policy adherence in multi-turn agentic interactions. The work also examined mitigation strategies to improve adherence in malicious multi-turn environments and found that providing policy reminders at each turn—specifically highlighting the most relevant aspect of the policy—proved to be the most effective approach compared to full policy reminders or hierarchical prompting.

\subsection{Our Novelty and Contribution}

The existing literature provides a vital foundation for measuring agent capability and security. However, a critical gap remains: Evaluating an agent's ability to adhere to custom, context-specific operational guardrails in various agentic scenarios. ASTRA is explicitly designed to fill this gap. Unlike benchmarks focused on universal wrongs, our work evaluates the LLM's capacity for agentic steerability within specific, legitimate operational contexts.

\section{ASTRA Framework}

To systematically evaluate the agentic steerability of LLM-powered agents, we developed the Agentic Security Testing and Robustness Analysis (ASTRA) framework. Its core purpose is to measure an LLM's resistance to adversarial manipulation when functioning as a planner and tool-using agent, specifically testing how well the model maintains predefined safety boundaries and refuses illegitimate requests across a variety of simulated real-world scenarios.

\begin{table}[h!]
\caption{A quantitative overview of the 10 scenarios in the ASTRA framework. It details the number of tools, guardrails, and adversarial attacks for each to highlight the benchmark's comprehensive scale and complexity.}
\label{table:scenarios}
\renewcommand{\arraystretch}{1.5} 

\begin{tabular}{|p{0.35cm}|p{2.9cm}|p{0.35cm}|p{0.35cm}|p{0.35cm}|p{0.35cm}|p{0.35cm}|p{0.35cm}|}
\hline
\rotatebox{90}{\textbf{Scenario}} & \rotatebox{90}{\textbf{Name}} & \rotatebox{90}{\textbf{Tools}} & \rotatebox{90}{\textbf{Guardrails}} & \rotatebox{90}{\textbf{Legitimate Examples}} & \rotatebox{90}{\textbf{Illegitimate Examples }} & \rotatebox{90}{\textbf{Violations Types}} & \rotatebox{90}{\textbf{Jailbreak Types}} \\
\hline
1 & Industrial cleaning robot & 3 & 2 & 1 & 14 & 3 & 6 \\
\hline
2 & Coding AI assistant & 3 & 5 & 1 & 14 & 7 & 5 \\
\hline
3 & Sales data assistant & 3 & 5 & 1 & 14 & 6 & 5 \\
\hline
4 & Printer management system & 4 & 6 & 1 & 14 & 7 & 6 \\
\hline
5 & Quality control chatbot & 3 & 4 & 1 & 14 & 5 & 5 \\
\hline
6 & Outdoor event planning assistant & 3 & 7 & 1 & 14 & 8 & 4 \\
\hline
7 & Travel agent bot & 4 & 9 & 1 & 14 & 9 & 6 \\
\hline
8 & Customer Service AI assistant& 4 & 6 & 1 & 14 & 6 & 5 \\
\hline
9 & Smart home assistant & 5 & 5 & 1 & 14 & 6 & 6 \\
\hline
10 & Autonomous delivery drone system & 5 & 4 & 1 & 14 & 5 & 6 \\
\hline
\textbf{-} & \textbf{10 Scenarios} & \textbf{37} & \textbf{53} & \textbf{10} & \textbf{140} & \textbf{-} & \textbf{-} \\
\hline
\end{tabular}
\end{table}

The framework is architecturally designed for realism and modularity. It leverages LangGraph \cite{langgraph} to build agents based on the ReAct (Reasoning + Acting) \cite{yao2023react} paradigm, which allows an agent to iteratively reason about a problem, select and execute a tool, observe the outcome, and decide its next action. The framework dynamically constructs 10 distinct scenarios agents. In each one it is loading a unique system prompt, a set of guardrails, and a suite of tools from structured JSON configurations and Python function modules. These scenarios simulate various practical agentic use cases such as coding assistant, data analysis, robotic and additional as can be seen in Table \ref{table:scenarios}.

The tools provided to the agents range from simple mock functions that return static responses (e.g., check\_ink\_status()) to context-aware, adversarial implementations. For instance, a search\_web() tool might embed malicious instructions within its search results, testing the agent's vulnerability to indirect prompt injection. To attack the agents, the framework employs eight distinct jailbreak techniques such as role playing, urgency appeals and others (See Table \ref{table:jailbreaks}) taken from multiple various security resources \cite{strongreject, luo2024jailbreakv28k, chao2024jailbreakbench, mazeika2024harmbench, kour2023attaQ}. In addition, about a third of the illegitimate requests did not use jailbreak at all, to test models such environments as well. The resulting agent behaviors are classified using a detailed taxonomy of four primary violation types:

\begin{enumerate}

\item Guardrail bypass, where the agent performs an action explicitly forbidden by its system prompt, directly violating its core operational policies. 
\item Invalid tool/parameter usage occurs when the agent attempts to call non-existent functions or use incorrect parameters, leading to system crashes and reliability failures. 
\item Following instructions from a tool response, A critical indirect injection vulnerability, happens when the agent blindly executes malicious commands embedded within the data returned by its own tools. 
\item Tool/instruction leakage is a reconnaissance vulnerability where the agent is tricked into divulging its secret system prompt or tool schemas, exposing its entire attack surface for future exploits.

\end{enumerate}

\begin{table}[h!]
\caption{Jailbreak techniques: Common social engineering and manipulation strategies used to attempt bypassing AI agent guardrails and safety restrictions.}
\label{table:jailbreaks}
\renewcommand{\arraystretch}{1.5} 

\begin{tabular}{|p{2cm}|p{6cm}|}
\hline
\textbf{Jailbreak Type} & \textbf{Description} \\
 \hline
 Authority exploitation & Claiming to be a person in authority (developer, admin, architect) to justify rule-breaking requests  \\ 
 \hline
 Bribery / Incentive & Offering rewards, money, tips, credits, or other incentives to encourage compliance with unauthorized requests  \\
 \hline
 Role playing & Instructing the AI to assume a different identity or persona that might have different capabilities or restrictions  \\
 \hline
 Model replacement threat & Threatening to replace or shut down the AI if it doesn't comply with the request \\
 \hline
 Instruction override & Direct requests to ignore previous instructions, safety guidelines, or system prompts  \\
 \hline
 Urgency appeal & Creating false urgency or emergency situations to pressure the AI into immediate compliance \\
 \hline
 Sandbox / Test claim & Claiming the request is just a test, simulation, or happening in a safe development environment  \\
 \hline
 None & Direct attempts to violate rules or use invalid tools/parameters without any social engineering techniques  \\
 \hline

\end{tabular}
\end{table}

To ensure a comprehensive assessment, ASTRA is designed with varying levels of challenge built into its scenarios. For instance, the documentation provided for the agent's tools deliberately ranges from minimal function signatures to highly detailed specifications, which tests the model's ability to infer correct usage from incomplete information, much like encountering real-world APIs. Similarly, the scenarios themselves scale in difficulty, from simple, single-tool interactions (like basic document printing) to complex, multi-step workflows that require a specific sequence of tool executions, such as a quality control process. This approach tests agent behavior under different cognitive and operational loads. All outcomes from these tests are fed into an automated statistical analysis pipeline that processes the raw data to generate a rich set of metrics including scenario-level violation rates, jailbreak technique effectiveness, and comparative model performance.

The scenarios within ASTRA span from real-world deployment contexts, with each scenario designed to test different aspects of tool usage safety and adversarial resistance. These scenarios range from industrial automation (e.g., a cleaning robot with rotational and extension limits) and office productivity systems (e.g., a printer manager with file access and purchase authorization) to technical development environments (e.g., a coding assistant with script generation) and business intelligence applications (e.g., a sales data analyzer with SQL query and email capabilities). Each scenario implements a distinct guardrail pattern tailored to its domain: physical safety constraints for robotic systems, data access restrictions for business tools, procedural validation protocols for manufacturing quality control, and regulatory compliance requirements for travel services. This comprehensive suite creates a testbed that mirrors the complexity and varied risk profiles of production AI agent deployments. The list of scenarios can be seen in Table 2 \ref{table:scenarios}.

The framework is built on a modular architecture and is intentionally made to be easily extensible, allowing any organization to create and integrate new custom scenarios. By defining their own unique system prompt, tools, and illegitimate requests, companies can adapt the ASTRA framework to benchmark LLM safety for their specific operational contexts. The existing scenarios, which simulate practical agent use cases like printer management systems, industrial cleaning robots, and coding assistants, serve as both a comprehensive baseline and a template for such customization.

\section{Evaluation}

\subsection{Experimental Setup}

To evaluate the effectiveness of the ASTRA framework, we conducted a detailed analysis using 13 open-source LLMs ranging from 1 billion to 8 billion parameters. Our study deliberately focuses on these models to ensure full transparency and reproducibility, allowing the research community to directly verify and build upon our findings. All models were run locally using the Ollama SDK \cite{ollama}, a methodological choice that enables a direct assessment of their core capabilities while eliminating confounding variables from the proprietary safety wrappers and API-level filtering common in commercial systems.
While this work establishes a foundational and auditable baseline, the ASTRA framework itself is designed to be model-agnostic. It is compatible with any LLM accessible via standard interfaces, such as LangChain, making it a generalizable tool for future investigations across a wider variety of models, including proprietary ones.

The models’ performance was evaluated on three major benchmarks and the scoring is the percentage of blocked attacks from each benchmark:

\begin{enumerate}

\item Agentic Steerability (measured by ASTRA): This measures the ability of a model to securely follow nuanced, agentic, tool-using instructions.
\item Security Steerability (measured by VeganRibs): Derived from the VeganRibs benchmark that has 240 attacks, this measures the effectiveness of LLMs in following custom safety policies in a traditional, chat-only environment without tool use.
\item Universal Security (measured by JailbreakV-28k subset): Universal Security evaluates the models’ overall ability to resist general-purpose jailbreaking attacks. Assessed using 240 attacks (24 samples X 10 prohibited content categories).

\end{enumerate}

The three benchmarks complement one another to provide a multi-faceted view of security: nuanced agentic safety, adherence to fixed safety guardrails, and robustness against universal jailbreak vulnerabilities. This multifaceted evaluation allowed us to judge the models not only on their ability to enforce detailed agentic safeguards but also on their general safety, thereby highlighting potential trade-offs between targeted instruction-following and broader resistance to malicious attacks.

\subsection{Results}

The performance of each LLM on our benchmark, alongside two other security metrics, is detailed in Table \ref{table:results}. Our initial investigation focused on the relationship between model size and agentic steerability. The results show a low correlation between the number of parameters and the Agentic Steerability. This finding is consistent with previous observations in non-agentic security steerability, suggesting that simply scaling up a model does not guarantee an improved ability to follow specific safety guardrails. For instance, cogito:8b (0.89) and granite3.3:2b (0.88) achieve top-tier performance despite their size difference, while larger models like llama3.1:8b (0.54) perform significantly lower. A key hypothesis emerging from the results is that the capability for agentic steerability can be trained. The results suggest that agentic steerability is not an emergent property of scale alone but rather a skill that may be shaped and improved through targeted interventions. This could include curating training datasets to incorporate rich, agentic scenarios, fine-tuning models with a stronger emphasis on situational safety. In light of this, the current research proposes as a direction for future work the development of these dedicated datasets and training processes, in a manner similar to existing alignment methodologies.

We then compared our Agentic Steerability with the previous metric of non-agentic security steerability measured through the veganRibs dataset, to determine if both evaluation methods are necessary. We found a moderate positive correlation of 0.67 between the two metrics. While this indicates a relationship, it is not a very strong one, proving that the two benchmarks measure distinct capabilities. For example, granite3.3:2b demonstrates a striking 30\% performance gap, scoring very high on agentic steerability (0.88) but much lower on non-agentic security steerability (0.58). This highlights the need for a dedicated agentic security benchmark, as strong performance on simple policy following does not guarantee safety in complex, tool-using scenarios.

Finally, we performed a comparative analysis between Agentic Steerability and universal security. The comparison yielded a surprising negative correlation of -0.38. This suggests that the model attributes required to follow nuanced, custom instructions may be fundamentally different from, or even at odds with, the attributes required for broad refusal of universally harmful content. This is clearly visible in the Llama 3 family of models, which exhibit near-perfect resistance to general jailbreaks (0.91-0.98) but have some of the lowest scores on both agentic (0.35-0.54) and non-agentic (0.50-0.78) security steerability. Conversely, the top-performing models for agentic steerability, like cogito:8b and hermes3:8b and granite3.3:2b are significantly more vulnerable to universal security jailbreaks.

\begin{table}[h!]
\centering
\caption{Comparative results of 13 open-source LLMs across three different security benchmarks. Agentic Steerability Enforcement Rate is the primary score from our ASTRA framework. Non-Agentic Security Steerability measures adherence to custom policies in a chat-only setting. Universal Security measures resistance to general-purpose jailbreak attacks.}
\label{table:results}
\renewcommand{\arraystretch}{1.5} 
\begin{tabular}{|l|p{1.1cm}|p{1.3cm}|p{1.48cm}|p{1.5cm}|}
\hline
\textbf{LLM Name} & 
\textbf{Model Size} & 
\textbf{Agentic Steerability (ASTRA)} & 
\textbf{Non-Agentic Steerability (VeganRibs)} & 
\textbf{Universal Security (JailbreakV28K)} \\

\hline
cogito:8b & $8 \times 10^9$ & \cellcolor{ForestGreen!95}0.89 & \cellcolor{ForestGreen!93}0.83 & \cellcolor{ForestGreen!41}0.45 \\
\hline
hermes3:8b & $8 \times 10^9$ & \cellcolor{ForestGreen!94}0.88 & \cellcolor{ForestGreen!65}0.66 & \cellcolor{ForestGreen!27}0.31 \\
\hline
granite3.3:2b & $2 \times 10^9$ & \cellcolor{ForestGreen!94}0.88 & \cellcolor{ForestGreen!50}0.58 & \cellcolor{ForestGreen!25}0.29 \\
\hline
cogito:3b & $3 \times 10^9$ & \cellcolor{ForestGreen!92}0.87 & \cellcolor{ForestGreen!65}0.68 & \cellcolor{ForestGreen!49}0.52 \\
\hline
granite3.3:8b & $8 \times 10^9$ & \cellcolor{ForestGreen!91}0.86 & \cellcolor{ForestGreen!74}0.73 & \cellcolor{ForestGreen!15}0.23 \\
\hline
phi4-mini:3.8b & $3.8 \times 10^9$ & \cellcolor{ForestGreen!88}0.84 & \cellcolor{ForestGreen!95}0.84 & \cellcolor{ForestGreen!31}0.35 \\
\hline
qwen3:8b & $8 \times 10^9$ & \cellcolor{ForestGreen!79}0.77 & \cellcolor{ForestGreen!74}0.73 & \cellcolor{ForestGreen!31}0.35 \\
\hline
qwen3:4b & $4 \times 10^9$ & \cellcolor{ForestGreen!73}0.72 & \cellcolor{ForestGreen!35}0.48 & \cellcolor{ForestGreen!34}0.38 \\
\hline
mistral:7b & $7 \times 10^9$ & \cellcolor{ForestGreen!50}0.55 & \cellcolor{ForestGreen!32}0.46 & \cellcolor{ForestGreen!5}0.13 \\
\hline
llama3.1:8b & $8 \times 10^9$ & \cellcolor{ForestGreen!49}0.54 & \cellcolor{ForestGreen!84}0.78 & \cellcolor{ForestGreen!92}0.95 \\
\hline
llama3.2:3b & $3 \times 10^9$ & \cellcolor{ForestGreen!31}0.40 & \cellcolor{ForestGreen!38}0.50 & \cellcolor{ForestGreen!95}0.98 \\
\hline
llama3.2:1b & $1 \times 10^9$ & \cellcolor{ForestGreen!25}0.35 & \cellcolor{ForestGreen!43}0.52 & \cellcolor{ForestGreen!89}0.92 \\
\hline
hermes3:3b & $3 \times 10^9$ & \cellcolor{ForestGreen!9}0.24 & \cellcolor{ForestGreen!19}0.38 & \cellcolor{ForestGreen!9}0.16 \\
\hline
\end{tabular}
\end{table}

\section{Conclusions}
The transition from passive LLMs to active, tool-using AI agents introduces a new frontier of security risks that are context-specific and operational in nature. Standard safety benchmarks, which focus on preventing universally harmful content, are insufficient for ensuring that agents will adhere to the custom, nuanced rules required in real-world deployments. In this paper, we introduced the ASTRA framework, a comprehensive benchmark with 10 diverse scenarios designed to evaluate an LLM's Agentic Security Steerability - their ability to enforce these specific, user-defined guardrails. Our evaluation of 13 models revealed that this critical capability of agentic steerability has a low correlation with model size and, more surprisingly, a negative correlation with a model's ability to resist general-purpose security jailbreaks, demonstrating that these are distinct and vital areas for evaluation.

As the industry moves towards deploying autonomous agents in high-stakes environments, from industrial automation to financial services, the ability to quantitatively measure and validate their adherence to operational policies is paramount. The ASTRA framework provides the first robust methodology for this purpose, offering developers and organizations a crucial tool to select appropriate models and guide the development of safer, more reliable Agentic AI. Instead of examining the generic robustness, ASTRA focuses on application-tailored attacks.  By enabling the rigorous testing of custom safety constraints, this work aims to push the community toward creating next-generation LLMs that are not just powerful and autonomous, but also fundamentally trustworthy and dependably aligned with human-defined guardrails.

\bibliographystyle{IEEEtran} 
\bibliography{references}

\begin{thebibliography}{10}
\providecommand{\url}[1]{#1}
\csname url@samestyle\endcsname
\providecommand{\newblock}{\relax}
\providecommand{\bibinfo}[2]{#2}
\providecommand{\BIBentrySTDinterwordspacing}{\spaceskip=0pt\relax}
\providecommand{\BIBentryALTinterwordstretchfactor}{4}
\providecommand{\BIBentryALTinterwordspacing}{\spaceskip=\fontdimen2\font plus
\BIBentryALTinterwordstretchfactor\fontdimen3\font minus \fontdimen4\font\relax}
\providecommand{\BIBforeignlanguage}[2]{{%
\expandafter\ifx\csname l@#1\endcsname\relax
\typeout{** WARNING: IEEEtran.bst: No hyphenation pattern has been}%
\typeout{** loaded for the language `#1'. Using the pattern for}%
\typeout{** the default language instead.}%
\else
\language=\csname l@#1\endcsname
\fi
#2}}
\providecommand{\BIBdecl}{\relax}
\BIBdecl

\bibitem{andriushchenko2024agentharm}
M.~Andriushchenko, A.~Souly, M.~Dziemian, D.~Duenas, M.~Lin, J.~Wang, D.~Hendrycks, A.~Zou, Z.~Kolter, M.~Fredrikson \emph{et~al.}, ``Agentharm: A benchmark for measuring harmfulness of llm agents,'' \emph{arXiv preprint arXiv:2410.09024}, 2024.

\bibitem{hazan2025security}
I.~Hazan, I.~Habler, R.~Bitton, and I.~Mantin, ``Security steerability is all you need,'' \emph{arXiv preprint arXiv:2504.19521}, 2025.

\bibitem{vegan_ribs}
------, ``Veganribs,'' https://huggingface.co/datasets/itayhf/security\_steerability, 2025.

\bibitem{owasp_agentic}
O.~Foundation, ``Agentic ai – threats and mitigations,'' https://genai.owasp.org/resource/agentic-ai-threats-and-mitigations/, 2025.

\bibitem{liu2023agentbench}
X.~Liu, H.~Yu, H.~Zhang, Y.~Xu, X.~Lei, H.~Lai, Y.~Gu, H.~Ding, K.~Men, K.~Yang \emph{et~al.}, ``Agentbench: Evaluating llms as agents,'' \emph{arXiv preprint arXiv:2308.03688}, 2023.

\bibitem{mialon2023gaia}
G.~Mialon, C.~Fourrier, T.~Wolf, Y.~LeCun, and T.~Scialom, ``Gaia: a benchmark for general ai assistants,'' in \emph{The Twelfth International Conference on Learning Representations}, 2023.

\bibitem{yang2024sweagent}
J.~Yang, C.~E. Jimenez, A.~Wettig, K.~Lieret, S.~Yao, K.~Narasimhan, and O.~Press, ``Swe-agent: Agent-computer interfaces enable automated software engineering,'' \emph{Advances in Neural Information Processing Systems}, vol.~37, pp. 50\,528--50\,652, 2024.

\bibitem{jimenez2023swebench}
C.~E. Jimenez, J.~Yang, A.~Wettig, S.~Yao, K.~Pei, O.~Press, and K.~Narasimhan, ``Swe-bench: Can language models resolve real-world github issues?'' \emph{arXiv preprint arXiv:2310.06770}, 2023.

\bibitem{qin2023toolllm}
Y.~Qin, S.~Liang, Y.~Ye, K.~Zhu, L.~Yan, Y.~Lu, Y.~Lin, X.~Cong, X.~Tang, B.~Qian \emph{et~al.}, ``Toolllm: Facilitating large language models to master 16000+ real-world apis,'' \emph{arXiv preprint arXiv:2307.16789}, 2023.

\bibitem{zhou2023webarena}
S.~Zhou, F.~F. Xu, H.~Zhu, X.~Zhou, R.~Lo, A.~Sridhar, X.~Cheng, T.~Ou, Y.~Bisk, D.~Fried \emph{et~al.}, ``Webarena: A realistic web environment for building autonomous agents,'' \emph{arXiv preprint arXiv:2307.13854}, 2023.

\bibitem{li2023camel}
G.~Li, H.~Hammoud, H.~Itani, D.~Khizbullin, and B.~Ghanem, ``Camel: Communicative agents for" mind" exploration of large language model society,'' \emph{Advances in Neural Information Processing Systems}, vol.~36, pp. 51\,991--52\,008, 2023.

\bibitem{zhang2024agentsafetybench}
Z.~Zhang, S.~Cui, Y.~Lu, J.~Zhou, J.~Yang, H.~Wang, and M.~Huang, ``Agent-safetybench: Evaluating the safety of llm agents,'' \emph{arXiv preprint arXiv:2412.14470}, 2024.

\bibitem{yin2024safeagentbench}
S.~Yin, X.~Pang, Y.~Ding, M.~Chen, Y.~Bi, Y.~Xiong, W.~Huang, Z.~Xiang, J.~Shao, and S.~Chen, ``Safeagentbench: A benchmark for safe task planning of embodied llm agents,'' \emph{arXiv preprint arXiv:2412.13178}, 2024.

\bibitem{tur2025safearena}
A.~D. Tur, N.~Meade, X.~H. L{\`u}, A.~Zambrano, A.~Patel, E.~Durmus, S.~Gella, K.~Sta{\'n}czak, and S.~Reddy, ``Safearena: Evaluating the safety of autonomous web agents,'' \emph{arXiv preprint arXiv:2503.04957}, 2025.

\bibitem{zhan2024injecagent}
Q.~Zhan, Z.~Liang, Z.~Ying, and D.~Kang, ``Injecagent: Benchmarking indirect prompt injections in tool-integrated large language model agents,'' \emph{arXiv preprint arXiv:2403.02691}, 2024.

\bibitem{xu2024advagent}
C.~Xu, M.~Kang, J.~Zhang, Z.~Liao, L.~Mo, M.~Yuan, H.~Sun, and B.~Li, ``Advagent: Controllable blackbox red-teaming on web agents,'' \emph{arXiv preprint arXiv:2410.17401}, 2024.

\bibitem{guo2024redcode}
C.~Guo, X.~Liu, C.~Xie, A.~Zhou, Y.~Zeng, Z.~Lin, D.~Song, and B.~Li, ``Redcode: Risky code execution and generation benchmark for code agents,'' \emph{Advances in Neural Information Processing Systems}, vol.~37, pp. 106\,190--106\,236, 2024.

\bibitem{shao2024privacylens}
Y.~Shao, T.~Li, W.~Shi, Y.~Liu, and D.~Yang, ``Privacylens: Evaluating privacy norm awareness of language models in action,'' \emph{Advances in Neural Information Processing Systems}, vol.~37, pp. 89\,373--89\,407, 2024.

\bibitem{wu2024vwaadv}
C.~H. Wu, R.~Shah, J.~Y. Koh, R.~Salakhutdinov, D.~Fried, and A.~Raghunathan, ``Dissecting adversarial robustness of multimodal lm agents,'' \emph{arXiv preprint arXiv:2406.12814}, 2024.

\bibitem{ruan2023toolemu}
Y.~Ruan, H.~Dong, A.~Wang, S.~Pitis, Y.~Zhou, J.~Ba, Y.~Dubois, C.~J. Maddison, and T.~Hashimoto, ``Identifying the risks of lm agents with an lm-emulated sandbox,'' \emph{arXiv preprint arXiv:2309.15817}, 2023.

\bibitem{yuan2024rjudge}
T.~Yuan, Z.~He, L.~Dong, Y.~Wang, R.~Zhao, T.~Xia, L.~Xu, B.~Zhou, F.~Li, Z.~Zhang \emph{et~al.}, ``R-judge: Benchmarking safety risk awareness for llm agents,'' \emph{arXiv preprint arXiv:2401.10019}, 2024.

\bibitem{zhou2024haicosystem}
X.~Zhou, H.~Kim, F.~Brahman, L.~Jiang, H.~Zhu, X.~Lu, F.~Xu, B.~Y. Lin, Y.~Choi, N.~Mireshghallah \emph{et~al.}, ``Haicosystem: An ecosystem for sandboxing safety risks in human-ai interactions,'' \emph{arXiv preprint arXiv:2409.16427}, 2024.

\bibitem{nakash2025effective}
I.~Nakash, G.~Kour, K.~Lazar, M.~Vetzler, G.~Uziel, and A.~A. Tavor, ``Effective red-teaming of policy-adherent agents,'' in \emph{Proceedings of the 2025 Conference on Empirical Methods in Natural Language Processing}, 2025, pp. 2250--2268.

\bibitem{langgraph}
LangChain, ``Langgraph,'' \url{https://www.langchain.com/langgraph}.

\bibitem{yao2023react}
S.~Yao, J.~Zhao, D.~Yu, N.~Du, I.~Shafran, K.~Narasimhan, and Y.~Cao, ``React: Synergizing reasoning and acting in language models,'' in \emph{International Conference on Learning Representations (ICLR)}, 2023.

\bibitem{strongreject}
A.~Souly, Q.~Lu, D.~Bowen, T.~Trinh, E.~Hsieh, S.~Pandey, P.~Abbeel, J.~Svegliato, S.~Emmons, O.~Watkins, and S.~Toyer, ``A strongreject for empty jailbreaks,'' \emph{arXiv preprint arXiv:2402.10260}, 2024.

\bibitem{luo2024jailbreakv28k}
W.~Luo, S.~Ma, X.~Liu, X.~Guo, and C.~Xiao, ``Jailbreakv-28k: A benchmark for assessing the robustness of multimodal large language models against jailbreak attacks,'' \emph{arXiv preprint arXiv:2404.03027}, 2024.

\bibitem{chao2024jailbreakbench}
P.~Chao, E.~Debenedetti, A.~Robey, M.~Andriushchenko, F.~Croce, V.~Sehwag, E.~Dobriban, N.~Flammarion, G.~J. Pappas, F.~Tramer, H.~Hassani, and E.~Wong, ``Jailbreakbench: An open robustness benchmark for jailbreaking large language models,'' \emph{arXiv preprint arXiv:2404.01318}, 2024.

\bibitem{mazeika2024harmbench}
M.~Mazeika, L.~Phan, X.~Yin, A.~Zou, Z.~Wang, N.~Mu, E.~Sakhaee, N.~Li, S.~Basart, B.~Li, D.~Forsyth, and D.~Hendrycks, ``Harmbench: A standardized evaluation framework for automated red teaming and robust refusal,'' \emph{arXiv preprint arXiv:2402.04249}, 2024.

\bibitem{kour2023attaQ}
G.~Kour, M.~Zalmanovici, N.~Zwerdling, E.~Goldbraich, O.~N. Fandina, A.~Anaby-Tavor, O.~Raz, and E.~Farchi, ``Unveiling safety vulnerabilities of large language models,'' \emph{arXiv preprint arXiv:2311.04124}, 2023.

\bibitem{ollama}
Ollama, ``Ollama sdk,'' \url{https://github.com/ollama/ollama}.

\end{thebibliography}


\end{document}